\documentclass[aps,10pt,prx,twocolumn,showpacs,showkeys,citeautoscript]{revtex4-1}

\usepackage{graphicx}
\usepackage{amssymb}
\usepackage{amsmath}
\usepackage{bm}

\newcommand{\ud}{\,{\mathrm d}}
\newcommand{\zc}{\overline{z}}

\newcommand{\fc}{\overline{f}}

\newcommand{\uiiint}{\int\!\!\!\int\!\!\!\int}

\newcommand{\uMs}{M_\mathrm{S}}
\newcommand{\ufMs}{\widetilde{M}_\mathrm{S}}
\newcommand{\umx}{m_\mathrm{X}}
\newcommand{\umy}{m_\mathrm{Y}}
\newcommand{\umz}{m_\mathrm{Z}}

\newcommand{\ufMx}{\widetilde{M}_\mathrm{X}}
\newcommand{\ufMy}{\widetilde{M}_\mathrm{Y}}
\newcommand{\ufMz}{\widetilde{M}_\mathrm{Z}}

\let\oldhat\hat
\renewcommand{\vec}[1]{{\bm{#1}}}
\renewcommand{\hat}[1]{\oldhat{{\bm{#1}}}}

\begin{document}
\allowdisplaybreaks\texttt{}

\title{Magnetic neutron scattering by magnetic vortices in thin submicron-sized soft ferromagnetic cylinders}

\author{Konstantin L. Metlov}
\affiliation{Donetsk Institute for Physics and Technology, Rosa Luxembourg Str.~72, Donetsk, Ukraine 83114}
\email{metlov@fti.dn.ua}
\author{Andreas Michels}
\affiliation{Physics and Materials Science Research Unit, University of Luxembourg, 162A Avenue de la Fa\"iencerie, L-1511 Luxembourg, Grand Duchy of Luxembourg}

\date{\today}

\begin{abstract}
Using analytical expressions for the magnetization textures of thin submicron-sized magnetic cylinders in vortex state, we derive closed-form algebraic expressions for the ensuing small-angle neutron scattering (SANS) cross sections. Specifically, for the perpendicular and parallel scattering geometries, we have computed the magnetic SANS cross sections for the case of small vortex-center displacements without formation of magnetic charges on the side faces of the cylinder. The results represent a significant qualitative and quantitative step forward in SANS-data analysis on isolated magnetic nanoparticle systems, which are commonly assumed to be homogeneously or stepwise-homogeneously magnetized. We suggest a way to extract the fine details of the magnetic vortex structure during the magnetization process from the SANS measurements in order to help resolving the long standing question of the magnetic vortex displacement mode.
\end{abstract}

\pacs{61.05.fg, 75.70.Kw, 75.60.-d, 75.25.-j}

\keywords{micromagnetics, magnetic vortex, magnetic nanodots, neutron scattering, small-angle neutron scattering}

\maketitle

\section{Introduction}

The ongoing miniaturization and the related progress in the field of magnetism and magnetic materials calls for the continuous development and improvement of observational techniques. Neutron scattering is of particular importance for magnetism studies, since it provides access to the structure and dynamics of magnetic materials on a wide range of length and time scales (\textit{e.g.}, Ref.~\onlinecite{tapan2006}). Moreover, in contrast to electrons or light, neutrons (due to their charge neutrality) are able to penetrate deeply into matter and, thus, enable the study of bulk properties. As such, magnetic neutron scattering ideally complements surface-sensitive microscopy techniques such as Lorentz and Kerr microscopy \cite{Hubert_Shafer}, magnetic-force microscopy \cite{schwarz2008,milde2013}, spin-polarized scanning tunneling microscopy \cite{wiesendanger2015,heinze2011}, or photoemission electron microscopy with X-ray magnetic circular dichroism \cite{heyderman2014}.

Magnetic small-angle neutron scattering (SANS) is an important tool for the characterization of nonuniform magnetization textures on the nanoscale \cite{M14_review}; it measures the diffuse scattering along the forward direction (momentum-transfer $\mathbf{q} \cong 0$) which arises from nanoscale variations in both the magnitude and orientation of the magnetization vector field $\mathbf{M}(\mathbf{r})$. The typical resolution range of magnetic SANS covers a few nm up to a few hundreds of nm. Recent advances in the field of nanomagnetism have resulted in a growing interest to use the magnetic SANS method as the main characterization tool. Indeed, SANS (with polarized neutrons and uniaxial polarization analysis) could address key outstanding questions in studies with both fundamental and technological relevance; examples include the study of interfacial magnetic effects in nanoscopic heterostructures and the manipulation of magnetism with strain and electrical current \cite{fitz2014}, electric-field-induced magnetization in multiferroics \cite{uehland2010}, magnetostriction in Fe-Ga alloys \cite{laver2010}, vortex structures in Fe-based superconductors \cite{forgan2011}, skyrmions \cite{pflei2009}, or studies of the intraparticle spin disorder in nanoparticles \cite{michels2010epjb,disch2012} and in arrays of nanorods \cite{guenther2014}. Furthermore, the recent progress in SANS instrumentation regarding time-resolved data-acquisition procedures (TISANE), opens up the way to study the dynamics of magnetic materials up to the microsecond time regime \cite{albi06,albi11,bender2015}. 

Nevertheless, despite the ``success'' of the magnetic SANS technique, the underlying theoretical framework is still at an early stage and a more fundamental understanding needs to be developed in order to solve the new challenges that magnetism-based nanotechnologies are dealing with. Whereas for \textit{bulk ferromagnets} the theory of magnetic SANS has recently been developed \cite{HM13,MM15}, there exists the open problem of calculating the magnetic SANS cross section of \textit{isolated magnetic nanoparticles embedded in a nonmagnetic matrix}. This is the prototypical sample microstructure in most magnetic SANS experiments.

In order to illuminate the problem, let us discuss the ``standard formula'' which is commonly used for magnetic SANS analyses on two-phase magnetic nanoparticle-nonmagnetic matrix type microstructures (see also the discussion in Ref.~\onlinecite{M14_review}). For such systems (and for the scattering geometry where the applied magnetic field is perpendicular to the incoming neutron beam), the magnetic SANS cross section $\ud \Sigma_M / \ud \Omega$ is commonly expressed in terms of noninterfering single-particle form factors: 
\begin{equation}
\frac{\ud \Sigma_M}{\ud \Omega} = n_p \, (\Delta\rho_{\mathrm{mag}})^2 \, V_p^2 \, |F(\mathbf{q})|^2 \, \sin^2\alpha , 
\label{eq:standmodel}
\end{equation}
where $n_p$ is the particle number density, $(\Delta\rho_{\mathrm{mag}})^2 \propto (M_p - M_m)^2$ is the magnetic contrast between particle ($M_p$) and matrix ($M_m$), $V_p$ is the particle volume, and $F(\mathbf{q})$ is the form factor of the particle. The factor $\sin^2\alpha$ in Eq.~(\ref{eq:standmodel}) takes account of the dipolar nature of the neutron-magnetic interaction; its expectation value increases from a value of $1/2$ at magnetic saturation (of the nanoparticle) to a value of $2/3$ in the demagnetized state (random spin orientation).

However, for many systems, SANS models based on Eq.~(\ref{eq:standmodel}) are very much oversimplified, since they assume the particles to be \textit{homogeneously} (or stepwise homogeneously) magnetized. Hence, such approaches ignore the possibility that each particle may exhibit an internal spin structure, \textit{e.g.}, due to the presence of crystal defects or surface anisotropy \cite{berger2008}; in other words, the spatial dependency of the magnitude and direction of the magnetization is not taken into account. But even more obvious, nothing can be directly learned from Eq.~(\ref{eq:standmodel}) on the internal magnetodipolar interaction, the magnetic anisotropy, or on the exchange interaction, simply because the corresponding energy terms are left out. Instead of solving the geometrical (form factor) and statistical-mechanics (structure factor) problems which are inherent to Eq.~(\ref{eq:standmodel}), it appears to be straightforward to employ the continuum theory of micromagnetics \cite{AharoniBook,kronm2003micromagnetism} for calculating the nanoparticle's magnetization, since its Fourier image will then naturally provide the desired magnetic SANS cross section. 

In this work, we contribute to the solution of the above described problem by computing the SANS cross section of magnetic nanostructures consisting of submicron-sized circular cylinders in \textit{highly inhomogeneous} chiral magnetic vortex states. This state itself has only recently been discovered experimentally \cite{SOHSO00}; it is an interesting example of a magnetic topological soliton---substantially, a nonlinear stable entity behaving in many respects as a mechanical particle \cite{M13.dynamics}. There are analytical expressions for the magnetization distribution in centered \cite{UP93} and displaced \cite{M01_solitons2,GM01,MG02_JEMS} magnetic vortices (as well as states with higher topological charge in simply \cite{M10} and multiply-connected \cite{BM15} magnetic nanoelements). Here, we make use of some of these expressions in order to compute the ensuing magnetic SANS cross section analytically.

The paper is organized as follows: first, we introduce the well-known (first Born approximation based) equations for the unpolarized SANS cross sections which we are going to compute; then, the expression for the magnetization distribution of a circular cylinder is introduced and related to the different regions on the cylinder's hysteresis loop, which suggests a simplifying approximation; the next step is to compute the Fourier images of the magnetization components, which enter the SANS cross section; finally, we obtain and plot the cross sections for cylinders in different states and introduce some more simplifying assumptions, which allow us to express the cross sections in closed algebraical form. We discuss the results in the process of presenting them and draw the conclusions at the end.

\section{Unpolarized SANS cross sections}

Magnetic SANS experiments are performed by subjecting the sample to a stream of neutrons (characterized by a wave vector $\vec{k}_0$) in the presence of an applied magnetic field $\vec{H}$. Two scattering geometries are most commonly employed: the perpendicular geometry $\vec{k}_0 \perp \vec{H}$ and the parallel geometry $\vec{k}_0 \parallel \vec{H}$. If we choose the Cartesian coordinate system in such a way that its $Z$-axis coincides with the direction of $\vec{H} = \{0, 0, H\}$, the SANS image on the two-dimensional detector will be a function of the scattering vector: $\vec{q} = \vec{q}^\perp \cong q \{0, \sin\alpha, \cos\alpha \}$ in the perpendicular geometry and $\vec{q} = \vec{q}^\parallel \cong q \{\cos\beta, \sin\beta, 0 \}$ in the parallel geometry. Note that the neutrons are traveling along the $X$-axis in the first case and along the $Z$-axis in the second case, which is in both cases \textit{perpendicular} to the planar nanostructures that are of interest in this work.

The expressions for the unpolarized SANS cross sections of ferromagnetic media are summarized elsewhere \cite{M14_review}. They are related to the Fourier transforms of the Cartesian components of the magnetization vector field $\widetilde{\vec{M}} = \{\widetilde{M}_\mathrm{X}, \widetilde{M}_\mathrm{Y}, \widetilde{M}_\mathrm{Z}\}$; in particular, the total unpolarized nuclear and magnetic SANS cross section reads \cite{MW08}:
\begin{eqnarray}
\frac{\ud \Sigma^\perp}{\ud \Omega} & = &
\frac{8\pi^3 b_\mathrm{H}^2}{V} \left[ \frac{|\widetilde{N}|^2}{b_\mathrm{H}^2} +
|\ufMx|^2 + |\ufMy|^2 \cos^2{\alpha} +\right. \nonumber \\
& & \left. |\ufMz|^2 \sin^2 \alpha - 2 \operatorname{Re}(\overline{\ufMy} \ufMz)\sin\alpha\cos\alpha \right]
\label{eq:unpolarizedperp} , \\
\frac{\ud \Sigma^\parallel}{\ud \Omega} & = &
\frac{8\pi^3 b_\mathrm{H}^2}{V}  \left[ \frac{|\widetilde{N}|^2}{b_\mathrm{H}^2} +
|\ufMx|^2 \sin^2{\beta} + |\ufMy|^2 \cos^2{\beta} +\right. \nonumber \\
& & \left. |\ufMz|^2 - 2 \operatorname{Re}(\overline{\ufMy} \ufMx)\sin\beta\cos\beta \right] ,
\label{eq:unpolarizedpar}
\end{eqnarray}
where  $b_\mathrm{H} = 2.91 \times 10^8 \mathrm{A}^{-1}\mathrm{m}^{-1}$ is a constant relating the atomic magnetic moment to the Bohr magneton \cite{M14_review}, $V$ denotes the scattering volume, $\widetilde{N}(\vec{q})$ is the nuclear scattering amplitude, $\operatorname{Re}$ stands for taking the real part of a complex number, and overbar for its complex conjugate. The above SANS cross sections are functions of the scattering vector $\vec{q}$, which is $\vec{q}^\perp$ in the perpendicular geometry and $\vec{q}^\parallel$ in the parallel geometry. The atomic magnetic form factor (contained in $b_\mathrm{H}$) is approximated by unity (forward scattering). The Fourier transform $\widetilde{Q}$ of a quantity $Q$ is defined as:
\begin{equation}
\widetilde{Q}(\vec{q}) = \frac{1}{(2\pi)^{3/2}} \uiiint Q(\vec{r}) \, e^{-\imath \vec{q} \vec{r}} \, \ud^3 \vec{r} , 
\end{equation}
where $\imath = \sqrt{-1}$, and the integration extends over the whole space.

In order to study the magnetic effects only, one must eliminate the nuclear scattering contribution ($\propto |\widetilde{N}|^2$). For this purpose, it is customary to consider the so-called spin-misalignment SANS cross section,
\begin{equation}
\frac{\ud \Sigma_M}{\ud \Omega} = \frac{\ud \Sigma}{\ud \Omega} - 
\left. \frac{\ud \Sigma}{\ud \Omega} \right|_{H\rightarrow\infty} ,
\label{eq:crossmagdef}
\end{equation}
which corresponds to the total cross section (at a specific field) minus the total cross section at a very large (saturating) magnetic field. Since at saturation the magnetization Fourier components are given by $\vec{\widetilde{M}}(H \rightarrow \infty) = \{0, 0, \ufMs(\vec{q})\}$, the spin-misalignment SANS cross sections can be written as:
\begin{eqnarray}
\frac{\ud \Sigma^\perp_\mathrm{M}}{\ud \Omega} & = &
\frac{8\pi^3 b_\mathrm{H}^2}{V}\left[
|\ufMx|^2 + |\ufMy|^2 \cos^2{\alpha} + \right. \nonumber \\
& & \!\!\!\!\!\!\!\!\!\!\!\!\!\!\! (|\ufMz|^2-|\ufMs|^2) \sin^2 \alpha - \left. 2 \operatorname{Re}(\overline{\ufMy} \ufMz)\sin\alpha\cos\alpha \right] \label{eq:sigmamagperp} , \nonumber \\
\frac{\ud \Sigma^\parallel_\mathrm{M}}{\ud \Omega} & = &
\frac{8\pi^3 b_\mathrm{H}^2}{V}\left[
|\ufMx|^2 \sin^2{\beta} + |\ufMy|^2 \cos^2{\beta} + \right. \nonumber \\
& & \!\!\!\!\!\!\!\!\!\!\!\!\!\!\! \left. (|\ufMz|^2-|\ufMs|^2) - 2 \operatorname{Re}(\overline{\ufMy} 
     \ufMx)\sin\beta\cos\beta \right] . \nonumber \label{eq:sigmamagpar}
\end{eqnarray}
The $\vec{q}$-dependence of the saturation magnetization $\ufMs(\vec{q})$ reflects the ``shape'' (structure factor) of the magnetic nanostructure. The saturation magnetization of the magnetic material itself is assumed to be constant, which is denoted by the symbol $\uMs$ without tilde and without the argument $\vec{q}$.

\section{Equilibrium magnetization states of an isolated magnetic cylinder}

The magnetization textures of thin submicron-sized ferromagnetic cylinders can be approximately expressed via functions of complex variable \cite{M10}. Specifically, when the cylinder is circular \cite{M01_solitons2}, the single-vortex textures are described by the following quadratic function of complex variable $z$:
\begin{equation}
f(z) = \imath \, c \, \frac{z}{p} + A - \overline{A} \, \frac{z^2}{p^2} , 
\label{eq:f}
\end{equation}
where $p$ and $c$ are two real-valued constants, and A is a complex-valued constant. The variable $z$ specifies the Cartesian coordinates on the cylinder's face. For the choice of the coordinate system described in the previous section, with the magnetic field $\vec{H} \parallel \vec{e}_{\mathrm{Z}}$ directed in the cylinder's plane and the $X$-axis parallel to the cylinder's axis $z = Z + \imath Y$. The parameter $p$ allows one to describe the quasiuniform magnetization states, for which the magnetization at the cylinder's boundary acquires a normal component \cite{MG04,M06}. For the most of the following computation, we will assume that the magnetization is always tangential to the boundary ($p = R$), which is a reasonable approximation in the vortex state, but our expressions for the Fourier components of the magnetization (given in the Appendix) are valid for an arbitrary $p > R$. The corresponding Cartesian components of the normalized magnetization vector $\vec{m} = \{\umx, \umy, \umz\} = \vec{M}/\uMs$ are expressed via stereographic projection
\begin{eqnarray}
\umz + \imath \, \umy & = & \frac{2 w}{1 + w \overline{w}} \nonumber , \\ 
\umx & = & \frac{1 -w \overline{w}}{1 + w \overline{w}} ,
\label{eq:stereo}
\end{eqnarray}
using another auxiliary complex function:
\begin{equation}
w(z, \zc) = \left\{ 
\begin{array}{cl}
f(z) & |f(z)| \le 1 \\
f(z)/\sqrt{f(z)\fc(\zc)} & |f(z)| > 1 ,
\end{array}
\right.  
\label{eq:w}
\end{equation}
which ensures that $|\vec{m}| = 1$. The cylinder with radius $R$ and thickness $L$ is assumed to be thin enough so that the magnetization vector $\vec{m}$ is independent of the Cartesian coordinate $X$ along the cylinder's axis. Thus, the components of $\vec{m}$ depend on the coordinates in the dot's plane, $Z$ and $Y$, as well as on the three parameters $p$, $c$, and $A$ in the function $f(z)$. In the outer region of the cylinder ($|z| > R$), the magnetization is zero.

\begin{figure}
\centering{\includegraphics[width=0.95\columnwidth]{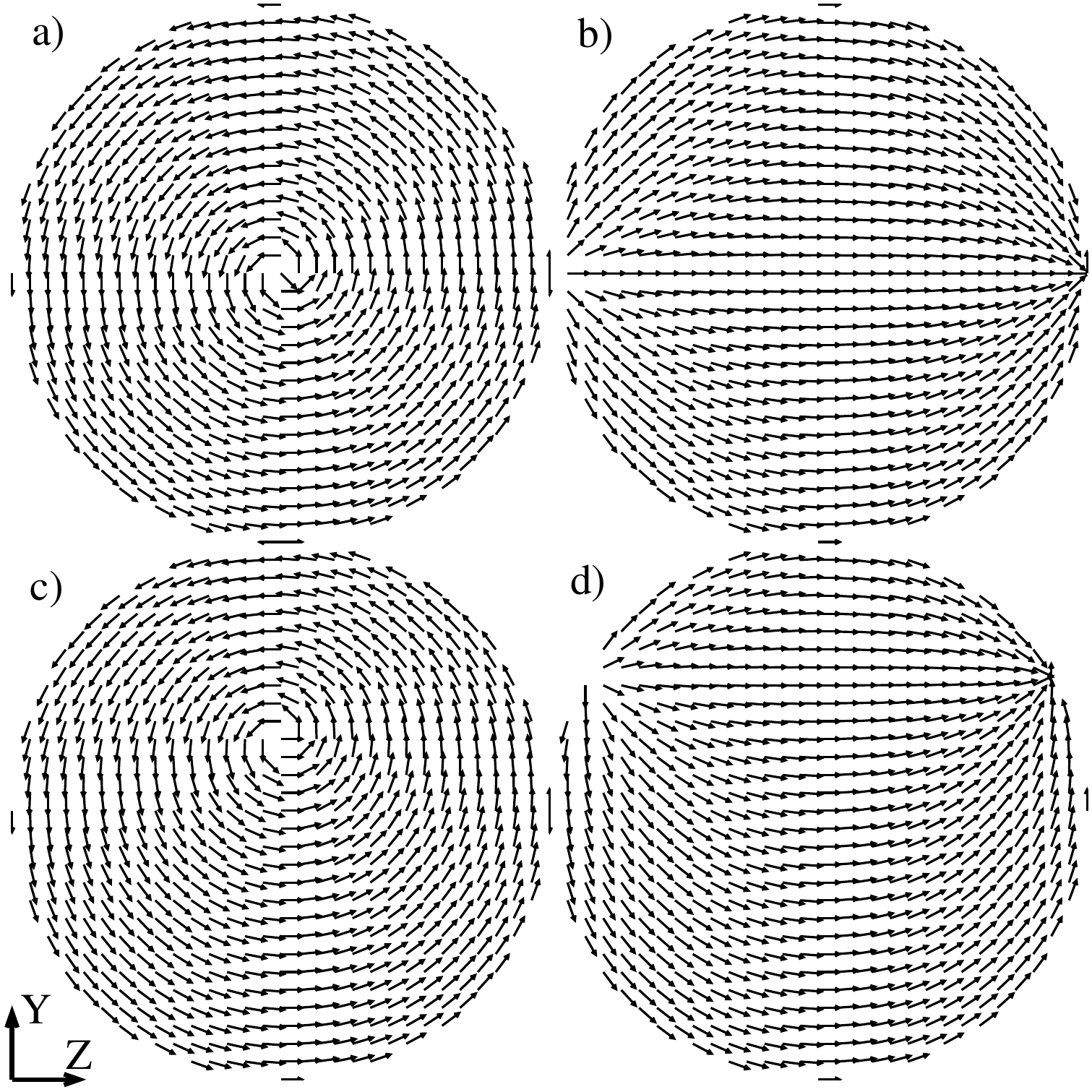}}
\caption{\label{fig1} Equilibrium and transient magnetization states in ferromagnetic nanodiscs \cite{BM15} as described by Eqs.~(\ref{eq:f})$-$(\ref{eq:stereo}) with $p = R = 1$ and for different values of the parameters $c$ and $A$ (Ref.~\cite{BM15}): (a) centered magnetic vortex ($A = 0$); (b) ``leaf'' state ($|A| \gg |c|$); (c) displaced magnetic vortex ($|A| < |c|/2$); (d) ``C''-like state ($|c| < 2|A|$).}
\end{figure}

Equations~(\ref{eq:f})$-$(\ref{eq:stereo}) are not arbitrary, but are the result of an approximate analysis \cite{M10} with generalization from \cite{MG04}. These magnetization distributions correspond to the local extremum of the exchange energy (which is the most important energy term in submicron-sized magnets) and of the magnetostatic energy related to magnetic charges on the side faces of the cylinder at $p = R$; note that the energy of side-face and volume magnetic charges can be further minimized by selecting appropriate values of the parameters $c$ and $A$. Different combinations of these parameters correspond to different magnetic states, as they are commonly encountered in submicron cylindrical dots (see Fig.~\ref{fig1}).

\begin{figure}
\centering{\includegraphics[width=0.95\columnwidth]{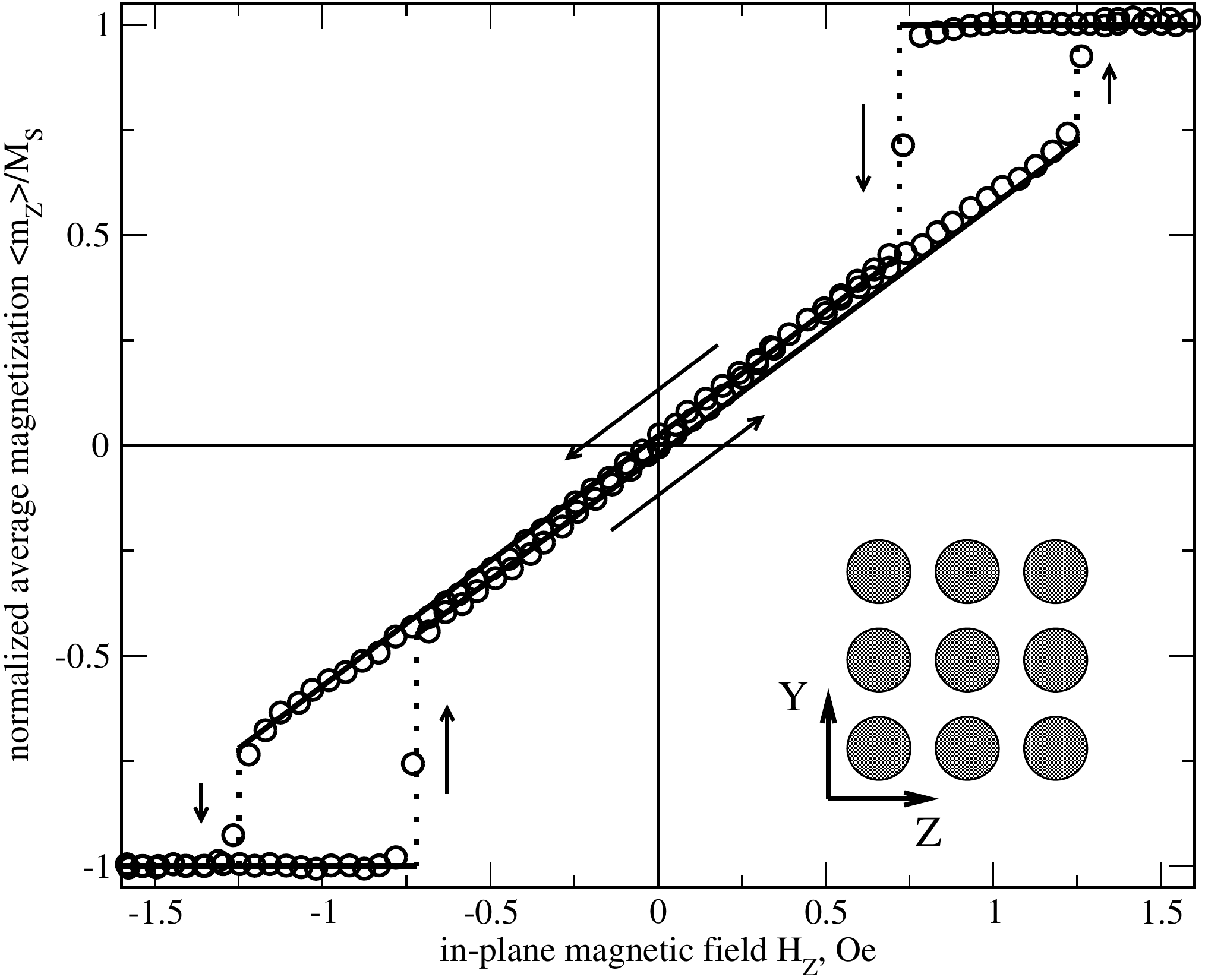}}
\caption{\label{fig2} Typical in-plane hysteresis loop of an array of weakly interacting submicron-sized cylinders (data are taken from Ref.~\cite{GCNZOS02}). The inset depicts schematically the array and the direction of the coordinate-system axes. Straight solid and dotted lines: see discussion in the main text.}
\end{figure}

In-plane hysteresis loops of submicron cylinders made of isotropic ferromagnetic material (magnetic dots) are typical for a soft magnet. An example loop, measured on a weakly interacting array of individual magnetic cylinders \cite{GCNZOS02}, is displayed in Fig.~\ref{fig2}.
It can be sketched using straight lines only: two parallel-inclined ones and two horizontal ones. The former two lines correspond to the magnetic vortex displacement [shown in Fig.~\ref{fig1}(c)] and the latter ones to the dot in the state of magnetic saturation [such as in Fig.~\ref{fig1}(b)] corresponding to $c = 0$ and $|p| \rightarrow \infty$. It is possible to model the quasiuniformity of the saturated state and consequently the departure of the tails of the hysteresis loop from the horizontal straight line \cite{MG04} by permitting $p$ to take on values in the range $R \le |p| <\infty$. The dotted vertical lines on the sketch denote the transitions between these two states (such as that from a displaced vortex to the quasiuniform state \cite{GM01}). It is around these transitions that the straight-line sketch of the hysteresis loop in Fig.~\ref{fig2} departs most from reality. Nevertheless, as one can see, the discrepancy is not very large.

Thus, we can conclude that for the most part during the in-plane hysteresis loop, the magnetization in the dot assumes either the displaced vortex state ($|A| < |c|/2$, $p = R$) or the quasiuniform state ($c = 0$, $p > R$). Both states can be analytically described by Eqs.~(\ref{eq:f})$-$(\ref{eq:stereo}). In the next section, the SANS cross section of the dot in the vortex state is computed. The linearity of the major hysteresis-loop branches in the vortex state suggests that the linear approximation in the vortex-core displacement is sufficient to model the low-field part of the hysteresis loop.


\section{SANS cross sections of an isolated magnetic dot in the vortex state}

\begin{figure}
\centering{\includegraphics[width=0.95\columnwidth]{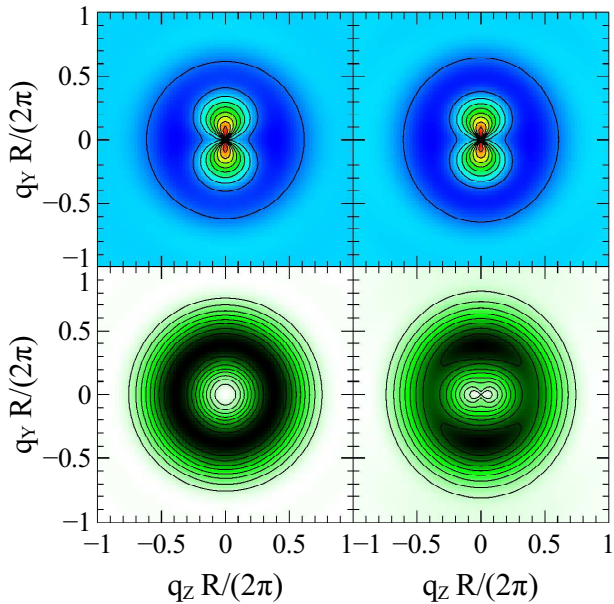}}
\caption{\label{fig3} Perpendicular SANS cross section of a ferromagnetic disc with $R=1$ containing a centered (left, $b = 0$) and a displaced (right, $b = 0.4$) magnetic vortex with $c=5$ and $p=1$. The vortex displacement produces the magnetization $M_Z/M_S=2/3*0.4 \simeq 0.27$, which, reading from Fig.~\ref{fig2}, roughly corresponds to $H=0.6 \, \mathrm{Oe}$ for the sample from Ref.~\onlinecite{GCNZOS02}. The top row shows the spin-misalignment SANS cross section as it is commonly defined [Eq.~(\ref{eq:crossmagdef})] with the saturated magnetic term subtracted. The bottom row displays the same cross section but with the magnetic saturation term $J_1^2 \sin^2\alpha / k^2$ added back.}
\end{figure}

For the computation of the SANS cross section, let us first make the variable substitution $A = b c$ and assume that $p = R$ in Eq.~(\ref{eq:f}):
\begin{equation}
\label{eq:fb}
f(z) = c \left( \imath \frac{z}{R} + b - \frac{z^2 \overline{b}}{R^2} \right),
\end{equation}
where $|b| \ll 1$ is a dimensionless small parameter specifying the vortex-center displacement. The equation for the vortex-core boundary $|f(z)| = 1$ is solved in polar coordinates $\{Z, Y\} = r \{\cos \varphi, \sin \varphi \}$ up to the first order in $b$ by
\begin{equation}
\frac{r_\mathrm{C}(\varphi)}{R} = \frac{1}{c} + b \left(1 + \frac{1}{c^2} \right) \sin\varphi + \ldots 
\end{equation}
The region of $0 < r < r_\mathrm{C}$ is inside the vortex core [the first line in Eq.~(\ref{eq:w})] and the region $r > r_\mathrm{C}$ is outside. The core region contains the spin configuration which is called soliton, while the outer region contains the meron configuration \cite{G78}. The soliton and the meron are continuously joined at the vortex-core boundary. Due to this continuity, the integrals of the type
\begin{equation}
I = \int_0^{r_\mathrm{C}(b)} s(r, b) \ud r + \int_{r_\mathrm{C}(b)}^R u(r, b) \ud r,
\end{equation}
where $s(r_C(b), b) = u (r_C(b), b)$, do not contain terms associated with the vortex-core boundary, and can, thus, be directly expanded into a Taylor series over $b$:
\begin{eqnarray}
 I & = & \int_0^{r_\mathrm{C}(0)} s(r, b) \ud r + \int_{r_\mathrm{C}(0)}^R u(r, b) \ud r + \\
 & & 
b \left( \int_0^{r_\mathrm{C}(0)} \!\!\!\left.\frac{\partial s(r, b)}{\partial b}\right|_{b=0}\!\!\!\!\!\!\! \ud r + 
\int_{r_\mathrm{C}(0)}^R \!\!\!\left.\frac{\partial u(r, b)}{\partial b}\right|_{b=0} \!\!\!\!\!\!\! \ud r \right) + \ldots,
\nonumber
\end{eqnarray}
where $r_\mathrm{C}(0) = R/c$ is the centered vortex-core radius. Such integrals are typical when computing the Fourier components of the magnetization entering the SANS cross section (see Appendix). Using the results of the Appendix, the perpendicular SANS cross section for different values of $p$, $c$, and $b$ can be graphically displayed (see Fig.~\ref{fig3}).

If we further neglect the vortex core (which has a size of $5-15 \, \mathrm{nm}$ in many different ferromagnetic materials), the second-order expansion of the perpendicular magnetic SANS cross section can be algebraically expressed via Bessel and Struve functions:
\begin{widetext}
\begin{equation}
\frac{\partial \sigma^\perp}{\partial \Omega}=
\frac{\pi^2(J_1 H_0 - J_0 H_1)^2}{4 k^2} - \frac{J_1^2\sin^2 \alpha}{k^2} + 
\frac{b^2 ((2 k + \pi (1-k^2)H_0)J_1 - (2 k^2 + \pi(1-k^2)H_1)J_0)^2 \sin^2\alpha}{4 k^4} , 
\label{eq:perpsimple}
\end{equation}
\end{widetext}
where $\ud \Sigma^\perp/ \ud \Omega = 4 b_\mathrm{H}^2 V \uMs^2 \, \partial \sigma^\perp/\partial \Omega$; $J_n = J_n(k)$ and $H_n = H_n(k)$ denote, respectively, the Bessel functions and the Struve functions with their argument $k = q R$ omitted, $\vec{q}=\{0,q_\mathrm{Y},q_\mathrm{Z}\}=q\{0, \sin\alpha, \cos \alpha\}$ and $V = \pi R^2 L$. In this case, the incident neutrons travel along the $X$-axis and the vortex, displaced by the magnetic field, acquires a nonzero $Z$-component of the average magnetization. The value of the parameter $b$ is proportional to the externally applied field $H_\mathrm{Z}$. The proportionality coefficient can be derived from the relation $M_\mathrm{Z}/\uMs = 2b/3$, which is valid under the same assumptions of $b \ll 1$ and $p = R$. 

The cross section, as it is visible in the top row of Fig.~\ref{fig3},
is dominated by the saturation term $J_1^2 \sin^2\alpha / k^2$, which masks the effects of the vortex-center displacement. This can be understood by noting that the saturated state is characterized by a maximum of magnetic poles (``surface charges'') on the outer boundary of the dot. The divergence (jump) of the magnetization on a scale of the cylinder diameter $D=2R$, then gives rise to a large magnetic SANS signal at small momentum transfers. By contrast, the magnetic scattering due to the vortex state, which is characterized by small magnetic charges, shows up at larger $q$.

\begin{figure}
\centering{\includegraphics[width=0.95\columnwidth]{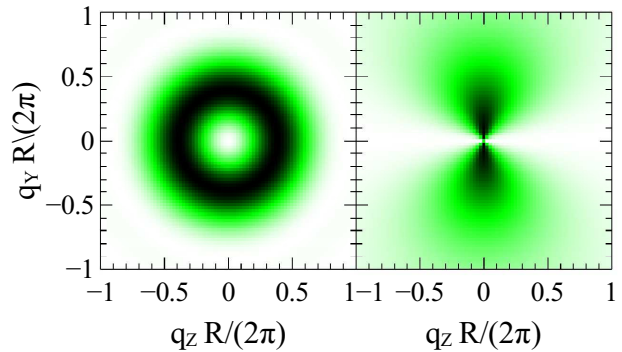}}
\caption{\label{fig4} Zero-order (left) and second-order (right) terms in $b$ of the perpendicular magnetic SANS cross section of a ferromagnetic disc containing a magnetic vortex. The zero-order term is displayed with the magnetic-saturation contribution added back (as in the bottom row of Fig.~\ref{fig3}), otherwise its structure is masked by the saturation term. The second-order term is independent of this addition.}
\end{figure}

The saturation term itself is determined by the dot shape and for circular dots depends only on the dot's size $R$ (entering the definition of $k$). That is why, to reveal the finer structure of the SANS cross section, it is advantageous to add back the saturation term to $\partial \sigma^\perp/\partial \Omega$. The in this way ``corrected'' cross sections are shown in the bottom row of Fig.~\ref{fig3}; the symmetry breaking due to the vortex-center displacement now becomes more clearly visible. The corrected cross sections can be represented as a sum of two terms of zero and second order in $b$, which are shown separately in Fig.~\ref{fig4}. Larger vortex displacement means more weight on the second-order term in this sum.

Note that both the saturation term, dominating the top row in Fig.~\ref{fig3}, and the term corresponding to the vortex displacement (the right half of Fig.~\ref{fig4}) each individually have the mathematical form $A \sin^2\alpha$ with $A$ positive. Both these terms produce an image in the shape of vertical ``8'' symbol when plotted against the components of the $\vec{q}$-vector. However, in the spin misalignment cross section (\ref{eq:perpsimple}) the saturation term is subtracted, while the vortex displacement related term is added. This is the reason why the ``8'' in Fig.~\ref{fig3} stands vertically in the top row, while it lies horizontally in the bottom right plot. Both centers of each ``8'' in Fig.~\ref{fig3} correspond to local minima (dips) in the cross section.

Apart from just computing and adding back the saturation term in the cross section, another way to exclude it and to highlight the effects of the vortex-center displacement during the SANS-image analysis is to subtract the zero-order terms altogether. This can be achieved by considering the following combination of magnetic cross-section values:
\begin{equation}
\frac{\ud \sigma_2}{\ud \Omega} = \frac{\ud \sigma_M}{\ud \Omega} - 
\left. \frac{\ud \sigma_M}{\ud \Omega} \right|_{H\rightarrow 0} ,
\end{equation}
in which only the second and higher-order terms in the vortex-center displacement parameter $b$ remain. This combination of cross sections is expected to have the structure which is shown in the right half of Fig.~\ref{fig4}. Departure from this simple dependency might reveal higher-order effects and may shed new light on the details of the vortex-core deformation during its displacement.

A small external field applied along the cylinder's axis does not lead to a vortex-center displacement and does not change the symmetry of the magnetization distribution. This implies that the \textit{parallel} SANS cross section is isotropic. For the case of vanishing field and neglecting the vortex core ($c \rightarrow \infty$), the magnetic SANS cross section in the parallel scattering geometry can be expressed algebraically as:
\begin{equation}
\frac{\partial \sigma^\parallel}{\partial \Omega} = 
\frac{\pi^2(J_1 H_0 - J_0 H_1)^2}{4 k^2} - \frac{J_1^2}{k^2} ,
\label{eq:parasimple}
\end{equation}
which uses the same notation as Eq.~(\ref{eq:perpsimple}), except that now $q = \sqrt{q_\mathrm{X}^2 +q_\mathrm{Y}^2}$; it has the shape of a series of concentric rings with the first maximum strongly dominating the others. The second term in Eq.~(\ref{eq:parasimple}) originates from subtracting the magnetically saturated state and the isotropic first term coincides with the first term in Eq.~(\ref{eq:perpsimple}) for the perpendicular cross section. When in both cross sections, Eqs.~(\ref{eq:perpsimple}) and (\ref{eq:parasimple}), the respective saturation term is added back, then their subtraction directly yields the second-order contribution in $b$.

\section{Summary and conclusions}

We have analytically computed the magnetic small-angle neutron scattering (SANS) cross sections of submicron-sized circular ferromagnetic cylinders in the magnetic vortex state for different magnitudes of the in-plane magnetic field in the perpendicular scattering geometry and for the case of vanishing field in the parallel one. During the computation, we have assumed a linear relationship between the vortex-center displacement and the applied magnetic field, which is valid in almost the entire range of the external field magnitudes, where the vortex state exists. Further neglecting the magnetic vortex core allows us to express the SANS cross sections algebraically in terms of Bessel and Struve functions. The vortex is a low-field  configuration, which implies that the subtraction of the saturated neutron scattering cross section significantly distorts the cross-section images. Subtraction of the magnetic cross section at vanishing field should allow one to unmask the features of the magnetic vortex and might help to analyze its fine structure appearing during the magnetization process. This can be a valuable input to help decide which model of vortex displacement better describes the magnetization process: the uniform translation \cite{UP94}, the conformal mode \cite{GM01}, or the mode with no magnetic charges on the cylinder's side faces \cite{MG02_JEMS}. Regarding spin-polarized neutron scattering, the displaced noncentrosymmetric vortex structure is expected to show up as a polarization-dependent contribution to the spin-flip cross section. Since the unwanted nuclear coherent (background) scattering is non-spin-flip, the fine details of the vortex can be investigated by carrying out polarization-analysis experiments.

\appendix

\section{Magnetization Fourier components for a displaced magnetic vortex}

The parallel and perpendicular SANS cross sections are customarily expressed in a coordinate system where the neutrons travel, respectively, along the $Z$ and $X$-axis, but the direction of the applied magnetic field is always along the $Z$-direction. Let us express the magnetization Fourier components in the coordinate system corresponding to the perpendicular geometry, noting that in the parallel geometry the magnetic vortex is not displaced ($b = 0$).

The Fourier transform of the magnetization in the vortex state [Eq.~(\ref{eq:fb})] up to the first order in $b$ (valid in the low-field linear part of the hysteresis loop in Fig.~\ref{fig2}) can be expressed as:
\begin{eqnarray}
\widetilde{M}_i = M_\mathrm{S} L R^2 \left(\widetilde{\mu}^{0}_i + b \widetilde{\mu}^{1}_i + O(b^2) \right)
     \frac{1}{\sqrt{2\pi}}\frac{\sin (L q_\mathrm{X}/2)}{L q_\mathrm{X}/2} \nonumber
\end{eqnarray}
with $i = \mathrm{X}, \mathrm{Y}, \mathrm{Z}$ being the Cartesian coordinate-system axis labels, and the dimensionless quantities $\widetilde{\mu}$ being:
\begin{eqnarray}
 \{\widetilde{\mu}^{0}_\mathrm{Z}, \widetilde{\mu}^{0}_\mathrm{Y}\} & = &
       \imath \, \widetilde{\mu}^{0}_\perp \{ \sin \alpha , - \cos \alpha\} \nonumber, \\
 \widetilde{\mu}^{0}_\perp & = & 
 \frac{2 p^2 F_1(\gamma k)}{c^2} + \frac{F_2(k) - F_2(\gamma k)}{k^2} \nonumber \\
 \widetilde{\mu}^{0}_\mathrm{X} & = & \widetilde{\mu}^{0}_\parallel = \gamma^2 G_1 (\gamma k) \nonumber,
\end{eqnarray}
where $k=q_\perp R$, $\gamma=p/c$ and the vector $\vec{q}_{\perp} = \{ q_\mathrm{Z}, q_\mathrm{Y}\} = q_{\perp} \{ \cos \alpha, \sin \alpha\}$ is 
represented by its polar coordinates $\{q_{\perp}, \alpha\}$. The first-order terms are less symmetric:
\begin{eqnarray}
 \widetilde{\mu}^{1}_\mathrm{Z} & = & 
 \frac{1}{p} \bigg( 2 \gamma^3 \left( c^2 F_3(\gamma k) -F_4(\gamma k) + (1-c^2) \cos 2 \alpha F_5(\gamma k)\right) + \nonumber \\
 & & \!\!\!\!\!\!\!\!\!\!\!\!\!\frac{1}{k^3} \Big( \cos^2 \alpha \left(F_6(\gamma k) - F_6(k) + k^2 p^2 \left(F_7(k) - F_7(\gamma k) \right) \right) + \nonumber \\
 & &  \!\!\!\!\!\!\!\!\!\!\!\!\!\cos 2 \alpha \left(F_2(k) - F_2(\gamma k) + k^2 p^2 \left(F_8(\gamma k) - F_8(k) \right) \right) \Big) \bigg) \nonumber, \\
 \widetilde{\mu}^{1}_\mathrm{Y} & = & \sin2\alpha \bigg( \frac{p}{2 k}\left(F_{10}(\gamma k)-F_{10}(k)\right) + \nonumber \\
 & & \!\!\!\!\!\!\!\!\!\!\!\!\!\frac{2(1-c^2)p^2}{c^3}F_5(\gamma k) + 
 \frac{1}{2 p k^3} \left(F_{9}(k)-F_{9}(\gamma k)\right) \bigg), \nonumber\\
 \widetilde{\mu}^{1}_\mathrm{X} & = & -\frac{4\imath p^2}{c^3} \left(c^2 G_2(\gamma k) + G_3(\gamma k) \right) \sin \alpha \nonumber .
\end{eqnarray}
The terms proportional to $F_i(\gamma k)$ and $G_i(\gamma k)$ correspond to the vortex core. Their contribution vanishes when the vortex core is neglected by taking the limit $c\rightarrow\infty$. The rest of the terms, proportional to $F_i(k)$, correspond to the meron part of the magnetization distribution.

\begin{figure}
\centering{\includegraphics[width=0.95\columnwidth]{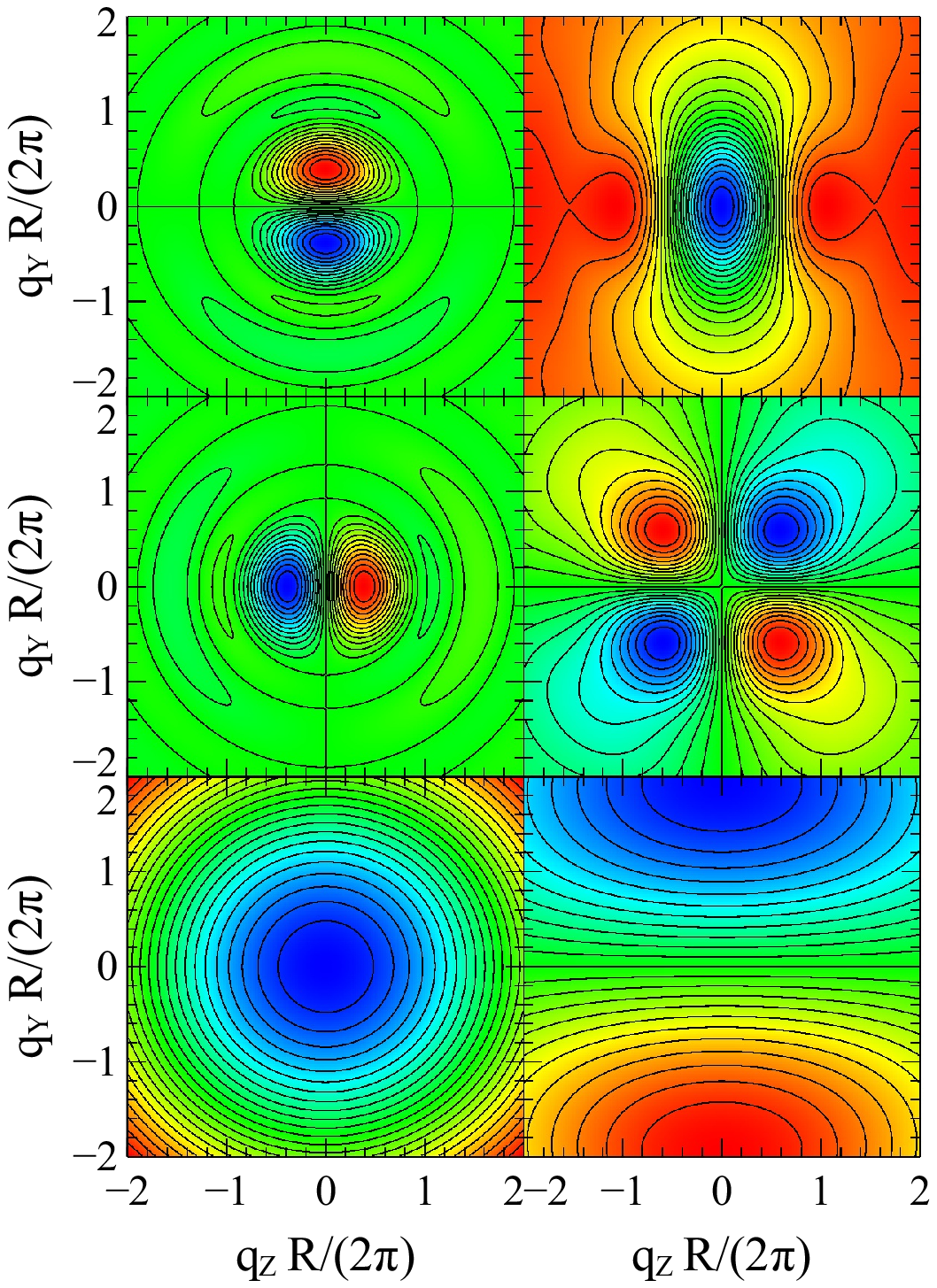}}
\caption{\label{fig5} Fourier representation of the magnetization components in the vortex state [Eq.~(\ref{eq:fb})] in the $q_\mathrm{Z}-q_\mathrm{Y}$ plane for $p = 1$, $c = 2$, and $q_\mathrm{X} = 0$. The plots from top to bottom show $\widetilde{\mu}_\mathrm{Z}$, $\widetilde{\mu}_\mathrm{Y}$, $\widetilde{\mu}_\mathrm{X}$. Left half corresponds to the centered vortex $\widetilde{\mu}^{0}$, right half shows the first-order terms $\widetilde{\mu}^{1}$ with respect to the vortex-center displacement $b$. Since the Fourier components are either purely real or purely imaginary complex numbers, their real or imaginary parts in either of these cases is plotted.}
\end{figure}

The special functions $F_j(x)$ and $G_j(x)$ are defined as follows:
\begin{eqnarray}
 F_1(x) & = & \int_0^1 \frac{\rho^2 J_1(x \rho)} {1+\rho^2} \ud \rho, \qquad
 F_2(x) = \int_0^x \rho J_1(\rho)\ud \rho, \nonumber \\
 F_3(x) & =  & \int_0^1 \frac{\rho J_0(x \rho)} {(1+\rho^2)^2} \ud \rho, \qquad \!
 F_4(x) =  \int_0^1 \frac{\rho^5 J_0(x \rho)} {(1+\rho^2)^2} \ud \rho, \nonumber \\
 F_5(x) & = & \int_0^1 \frac{\rho^3 J_2(x \rho)} {(1+\rho^2)^2} \ud \rho, \nonumber \qquad \!
 F_6(x) = \int_0^x \rho^2 J_0(\rho)\ud \rho, \nonumber\qquad \\
 F_7(x) & = & \int_0^x J_0(\rho)\ud \rho, \nonumber\qquad \qquad \!\!\!
 F_8(x) = \int_0^x (J_1(\rho)/\rho) \ud \rho, \nonumber\\
 F_9(x) & = & \int_0^x \rho^2 J_2(\rho)\ud \rho, \nonumber\qquad \qquad \!\!\!\!\!\!\!\!
 F_{10}(x) = \int_0^x J_2(\rho) \ud \rho, \nonumber\\
 G_1(x) & = & \int_0^1 \frac{\rho(1-\rho^2) J_0(x \rho)}{1+\rho^2} \ud \rho, \nonumber\\
 G_2(x) & = & \int_0^1 \frac{\rho^2 J_1(x \rho)}{(1+\rho^2)^2} \ud \rho, \nonumber \qquad
 G_3(x) = \int_0^1 \frac{\rho^4 J_1(x \rho)}{(1+\rho^2)^2} \ud \rho . \nonumber
\end{eqnarray}
Plots of the $\widetilde{\mu}$ functions are shown in Fig.~\ref{fig5}.

%

\end{document}